\documentclass[%
reprint,
superscriptaddress,
nofootinbib,
 amsmath,amssymb,
 aps,
 prd,
floatfix,
]{revtex4-2}

\usepackage{bm}% bold math
\usepackage{hyperref}% add hypertext capabilities
\hypersetup{
    colorlinks=true,
    linkcolor=blue,
    filecolor=blue,
    urlcolor=blue,
    citecolor=blue,
}

\usepackage[normalem]{ulem}
\usepackage{physics}
\usepackage{slashed}
\usepackage{mathrsfs}

\begin{document}

\title{Speed Variations
of Cosmic Photons and Neutrinos from Loop Quantum Gravity}\thanks{Published in Phys.~Lett.~B 836 (2023) 137613 \\ \url{https://doi.org/10.1016/j.physletb.2022.137613}}

\author{Hao Li}
 \affiliation{School of Physics,
 Peking University, Beijing 100871, China}

\author{Bo-Qiang Ma}%
 \email{mabq@pku.edu.cn}
\affiliation{School of Physics,
Peking University, Beijing 100871, China}

\affiliation{Center for High Energy Physics, Peking University, Beijing 100871, China} 

\affiliation{Collaborative Innovation Center of Quantum Matter, Beijing, China}

%\date{\today}

\begin{abstract}
Recently a series of analyses on the flight time of cosmic photons and neutrinos suggests that the speed of light \emph{in vacuo} takes the energy-dependent
form $v(E)\simeq 1-E/E_{\text{LIV}}^{\gamma }$ with
$E_{\text{LIV}}^{\gamma }\approx 3.6\times 10^{17}~\text{GeV}$, and meanwhile the speed of neutrinos is proposed to be $v(E)\simeq 1\pm E/E_{\text{LIV}}^{\nu }$ with $E_{\text{LIV}}^{\nu }\approx 6.5\times 10^{17}~\text{GeV}$ and $\pm {}$ representing
the helicity dependence. This novel picture immediately urges us to provide a satisfactory theoretical explanation. Among all the attempts to predict
the speed variations from quantum gravity, we find that loop quantum gravity can serve as a good candidate for explaining the aforementioned picture consistently.
\end{abstract}

%\keywords{Loop Quantum Gravity, Lorentz Invariance Violation, Photon Propagation, Neutrino Propagation}%Use showkeys class option if keyword display desired

\maketitle
Lorentz invariance, as a basic principle of relativity, is suspected
to break due to the quantum gravity~(QG) effect around the Planck scale
$E_{\text{Planck}}\approx 1.22\times 10^{19}~\text{GeV}{}$~(or the inverse
of the length scale $\ell _{P}\approx 1.6\times 10^{-33}~\text{cm}$). As
the footprint of the underlying QG effect, Lorentz invariance violation~(LIV)
is too minuscule to be directly observed in the laboratories, thus high
energy cosmic particles like photons and neutrinos might provide one of
the most promising opportunities to reveal the tiny LIV effects~\cite{Amelino-Camelia1998}.
In general, one expects that to the leading order of LIV, the modified
speed of a massless particle with energy $E$ can be expressed as
$v(E)\simeq c(1-\xi E)$, where $\xi{}$ is the LIV scale about the order
of magnitude of the Planck scale. Therefore given the precision accuracy
of current experimental facilities, the speed difference between two massless
particles with different energies can only be distinguishable when their
energies are high enough and the travel time differences are accumulated after a long
propagating distance. As a result, cosmic photons and neutrinos, especially
those from gamma-ray bursts~(GRBs) and active galactic nuclei~(AGNs) and
simultaneously energetic, open a unique window to the search of LIV from
their flights in the Universe~\cite{Amelino-Camelia1998}.

LIV researches utilizing high energy cosmic photons and neutrinos show
many interesting results, and we focus on one of them obtained by the analyses
of GRB photon travel time differences and possible GRB neutrino travel time
lags. The corresponding scenario to explain the data can be summarized
as
\begin{enumerate}
  \item For photons, we have
        $v(E)\simeq c(1-E/E_{\text{LIV}}^{\gamma })$ and
        $E_{\text{LIV}}^{\gamma }\approx 3.6\times 10^{17}~\text{GeV}{}$ without
        helicity dependence. This means that photons are all subluminal and high
        energy photons propagate slower than low energy photons. More details can
        be found in Refs.~\cite{Shao2010f,Zhang2015,Xu2016a,Xu2016,Amelino-Camelia:2016ohi,Amelino-Camelia:2017zva,Xu2018,Liu2018,Li2020,Zhu2021a,Chen2021}.
  \item For neutrinos and anti-neutrinos~(or vice versa), their speeds are
        $v(E)\simeq (1\mp E/E_{\text{LIV}}^{\nu })$ and
        $E_{\text{LIV}}^{\nu }\approx 6.5\times 10^{17}~\text{GeV}{}$. Consequently
        neutrinos and anti-neutrinos~(or Majorana neutrinos with opposite helicities)
        could be subluminal and superluminal respectively or visa versa, with the
        sign of the speed variations depending on the helicities. For more
        details one can refer to Refs.~\cite{Amelino-Camelia:2015nqa,Amelino-Camelia:2016fuh,Amelino-Camelia:2016ohi,Amelino-Camelia:2017zva,Huang:2018ham,Huang2019,Huang2022}.
\end{enumerate}
While full clarity on the suggested phenomenological picture must wait
for more data to verify, it makes sense to explore possible theoretical
frameworks that would provide support for the phenomenological picture
and establish whether by sharpening the phenomenology one could get insight
on some pure-theory proposals. Although there are various theoretical explanations
for only the photon sector or the neutrino sector, a combined explanation
of both photon and neutrino sectors is still desirable. To fill the gap
between phenomenological studies and theoretical constructions, we suggest
that loop quantum gravity~(LQG) is able to provide a viable approach to
understanding these phenomena consistently.

Loop quantum gravity~(LQG), one of the most promising theories aiming at
reconciling general relativity and quantum mechanics, has attracted a lot
of interest with numerous novel results. A comprehensive introduction to
LQG can be found in some excellent books such as~\cite{Rovelli:2004tv,Thiemann:2007pyv,Rovelli:2014ssa},
and for our purpose, we find it appropriate to only gather the essential
results as well as necessary details whenever they are needed, with more
information can be found in Refs.~\cite{Alfaro2002a,Alfaro:2002xz}. To
calculate LIV of photons and neutrinos from LQG, one considers a special
set of semiclassical states called the
\emph{would be semiclassical states}~(WBSCs). A WBSC $\ket{W}$ has a characteristic
length $\mathcal{L}$ which describes the discreteness of the spacetime
represented by this state. When the de Broglie wavelength
$\lambda{}$ of a particle satisfies $\ell_P\ll \lambda$ and the characteristic length
$\mathcal{L}$ satisfies $\ell _{P}\ll \mathcal{L}\le \lambda{}$, one can
compute the WBSC expectation of the corresponding Hamiltonian of the particle
and obtain an effective Hamiltonian from which we can read off the modified
dispersion relation and accordingly, the speed of that particle.

For the electromagnetic fields, the effective Hamiltonian results in the
following modified Maxwell equations~\cite{Alfaro2002a}:
%
%e1 #&#
\begin{eqnarray}
  A_{\gamma }(\grad \times \vec{B})&&-
  \frac{\partial \vec{E}}{\partial t}+ 2\ell _{P}^{2}\theta _{3}\grad ^{2}(
  \grad \times \vec{B})-2\theta _{8}\ell _{P}\grad ^{2}\vec{B}
  \nonumber
  \\
  &&+4\theta _{4}\mathcal{L}^{2}{\left (\frac{\mathcal{L}}{\ell _{P}}
  \right )}^{2\Upsilon _{\gamma}}\ell _{P}^{2}\grad \times ({\vec{B}}^{2}
  \vec{B})=0,
  \nonumber
  \\
  A_{\gamma }(\grad \times \vec{E})&&+
  \frac{\partial \vec{B}}{\partial t}+2\ell _{P}^{2}\theta _{3}\grad ^{2}(
  \grad \times \vec{E})
  \nonumber
  \\
  &&-2\theta _{8}\ell _{P}\grad ^{2}\vec{E}=0,
  \label{maxwell1}
\end{eqnarray}
with
%
%e2 #&#
\begin{equation}
  A_{\gamma}=1+\theta _{7}{\left (\frac{\ell _{P}}{\mathcal{L}}\right )}^{2+2
    \Upsilon _{\gamma}},
\end{equation}
and we supplement to Eq.~{\eqref{maxwell1}} the source-free conditions~\cite{Alfaro2002a}:
%
%e3 #&#
\begin{equation}
  \grad \cdot \vec{E}=\grad \cdot \vec{B}=0.
  \label{maxwell2}
\end{equation}
We leave the parameters $\theta _{i}$ (with $i=3,4,7,8$) and
$\Upsilon _{\gamma}$ to be determined by analyzing the phenomenological
picture later. However it should be noted that naturally we can expect
that all the $\theta _{i}$s are either of the order of magnitude of one
or are extremely close to zero. We neglect the non-linear part in Eq.~{\eqref{maxwell1}}
and assume that the solutions can be considered as superposition of plane
waves. Then substituting
\begin{equation*}
  \vec{E}\propto e^{i(\vec{k}\cdot \vec{x}-\omega t)},\qquad \vec{B}
  \propto e^{i(\vec{k}\cdot \vec{x}-\omega t)},\qquad k:=|\vec{k}|
\end{equation*}
into Eqs.~{\eqref{maxwell1}} and~{\eqref{maxwell2}}, we can get the modified
dispersion relation~\cite{Alfaro2002a}:
%
%e4 #&#
\begin{equation}
  \omega =k\left [1+\theta _{7}{\left (\frac{\ell _{P}}{\mathcal{L}}
      \right )}^{2+2\Upsilon _{\gamma}}-2\theta _{3}{(k\ell _{P})}^{2}\pm 2
    \theta _{8}(k\ell _{P})\right ],
  \label{photondispersion}
\end{equation}
which gives the speed of light from $v=\partial \omega /\partial k$~\cite{Alfaro2002a}:
%
%e5 #&#
\begin{equation}
  v_{\pm}^{\gamma}(k,\mathcal{L})\simeq 1\pm 4\theta _{8}(k\ell _{P})-6
  \theta _{3}{(k\ell _{P})}^{2}+\theta _{7}{\left (
    \frac{\ell _{P}}{\mathcal{L}}\right )}^{2\Upsilon _{\gamma}+2}.
  \label{photonspeed}
\end{equation}
This formula is valid only for small $\ell _{P}/\mathcal{L}$.

To compare Eq.~{\eqref{photonspeed}} with the aforementioned form of photon
speed, we first drop the third term which is of the second order of $\ell_P$ on the
right hand side. Then we adopt the mobile scale which relates
the characteristic scale $\mathcal{L}$ and the momentum $k$ by
$\mathcal{L}=k^{-1}$. The photon speed then can be written as
%
%e6 #&#
\begin{equation}
  v_{\pm}^{\gamma}(k)\simeq 1\pm 4\theta _{8}(\ell _{P} k)+\theta _{7}{(
    \ell _{P} k)}^{2\Upsilon _{\gamma}+2}.
  \label{photonspeed3}
\end{equation}
Obviously only the third term on the right hand side produces the helicity-independent
correction and we must set $\Upsilon _{\gamma}=-1/2$ such that our picture
can be reproduced qualitatively. Therefore we are led to consider the following
form:
%
%e7 #&#
\begin{equation}
  v_{\pm}^{\gamma}(k)\simeq 1+(\theta _{7}\pm 4\theta _{8})(\ell _{P} k).
\end{equation}
The $\theta _{8}$ term here still leads to birefringence effects and as
a result the total rotation angle between two oppositely polarized photons
with the same energy can be written as~\cite{Laurent2011,Toma2012,wei2021tests}
%
%e8 #&#
\begin{equation}
  \lvert \Delta \theta (E,z)\rvert \simeq
  \frac{2\theta _{8}\ell _{P} E^{2}}{H_{0}}\int _{0}^{z}
  \frac{(1+z^{\prime})\,\text{d}z^{\prime}}{\sqrt{\Omega _{m}{(1+z^{\prime})}^{3}+\Omega _{\Lambda}}},
\end{equation}
where $E$ is the observed energy, $z$ is the source redshift, and
$H_{0},\Omega _{m}$ and $\Omega _{\Lambda}{}$ are the cosmological constants~\cite{ParticleDataGroup:2014cgo}.
Therefore the best constraint on the rotation angle in the literature~\cite{wei2021tests}
can be transformed into a constraint on $\theta _{8}$:
$\theta _{8}\lesssim 10^{-16}$, for which we can simply let
$\theta _{8}$ be zero since photons usually have energies below
$10^{6}~\text{GeV}{}$ such that $\ell _{P} p\lesssim 10^{-13}$, and therefore
$\theta _{8}\ell _{P} p\sim{(\ell _{P} p)}^{2}$, and consequently we could
drop the term. As a result we have
%
%e9 #&#
\begin{equation}
  v_{\pm}^{\gamma}(k)\simeq 1+\theta _{7}\ell _{P} k,
\end{equation}
and
%
%e10 #&#
\begin{equation}
  \theta _{8}=0,\ \frac{1}{\theta _{7}\ell _{P}}\approx -3.6\times 10^{17}~
  \text{GeV}{},\ \text{and } \Upsilon _{\gamma}=-\frac{1}{2}.
  \label{photonparameters}
\end{equation}
That is to say, we have $\lvert \theta _{7}\rvert \approx 33.9$, which
is a reasonable numerical result near the order $\mathcal{O}(1)$.

On the other hand, for massless Majorana fermions, we have \cite{Alfaro:2002xz}
%
%e11 #&#
\begin{eqnarray}
  \left (i\frac{\partial}{\partial t}-i\hat{A}\vec{\sigma}\cdot \grad +
  \frac{\hat{C}}{2\mathcal{L}}\right )\xi (t,\vec{x})&&=0,
  \nonumber
  \\
  \left (i\frac{\partial}{\partial t}+i\hat{A}\vec{\sigma}\cdot \grad -
  \frac{\hat{C}}{2\mathcal{L}}\right )\chi (t,\vec{x})&&=0,
  \label{dirac}
\end{eqnarray}
where  $\sigma ^{i} $ with $i=1,2,3$ denote the standard Pauli matrices,
$\xi{}$ is a Weyl spinor,
$\chi (t,\vec{x})=i\sigma _{2}\xi ^{*}(t,\vec{x})$ and the two coefficients~(indeed
operators) are defined as~\cite{Alfaro:2002xz}
%
%e12 #&#
\begin{eqnarray}
  &&\hat{A}:=1+\kappa _{1}{\left (\frac{\ell _{P}}{\mathcal{L}}\right )}^{
    \Upsilon _{f}+1}+\kappa _{2}{\left (\frac{\ell _{P}}{\mathcal{L}}
    \right )}^{2\Upsilon _{f}+2}+\frac{\kappa _{3}}{2}\ell _{P}^{2}\grad ^{2},
  \nonumber
  \\
  &&\hat{C}/\hbar :=\kappa _{4}{\left (\frac{\ell _{P}}{\mathcal{L}}
    \right )}^{\Upsilon _{f}}+\kappa _{5}{\left (
    \frac{\ell _{P}}{\mathcal{L}}\right )}^{2\Upsilon _{f}+1}
  \nonumber
  \\
  &&{\ \ \ \ \ \ \ \ \ }+\kappa _{6}{\left (
    \frac{\ell _{P}}{\mathcal{L}}\right )}^{3\Upsilon _{f}+2}+
  \frac{\kappa _{7}}{2}{\left (\frac{\ell _{P}}{\mathcal{L}}\right )}^{
    \Upsilon _{f}}\ell _{P}^{2}\grad ^{2}.
\end{eqnarray}
Similarly, a reasonable assumption is that all the $\kappa _{i}$s are either
of the order of magnitude of one or just can be set to zero. Meanwhile
we assume $\kappa _{i}=0$ for $ i=4,5,6$ since these parameters merely
contribute to the renormalization of the mass which is of no interest in
this work~\cite{Alfaro:2002xz}. As a result, we use the simplified definition
of $\hat{C}$ in the following:
%
%e13 #&#
\begin{equation}
  \hat{C}=\frac{\hbar \kappa _{7}}{2}{\left (
    \frac{\ell _{P}}{\mathcal{L}}\right )}^{\Upsilon _{f}}\ell _{P}^{2}
  \grad ^{2}.
\end{equation}
According to Ref.~\cite{Alfaro:2002xz}, the dispersion relation can be
expressed as
%
%e14 #&#
\begin{equation}
  E_{\pm}=\sqrt{A_{f}^{2}p^{2}+{\left (\frac{C_{f}}{2\mathcal{L}}
  \right )}^{2}\pm B_{f}p},
  \label{fermiondispersion}
\end{equation}
where $p:=\lvert \vec{p}\rvert{}$ and
%
%e15 #&#
\begin{eqnarray}
  &&A_{f}=1+\kappa _{1}{\left (\frac{\ell _{P}}{\mathcal{L}}\right )}^{
    \Upsilon _{f}+1}+\kappa _{2}{\left (\frac{\ell _{P}}{\mathcal{L}}
    \right )}^{2\Upsilon _{f}+2}
  \nonumber
  \\
  &&{\ \ \ \ \ \ \ \ \ \ \ \ }-\frac{\kappa _{3}}{2}\ell _{P}^{2}p^{2},
  \nonumber
  \\
  &&B_{f}=A_{f}\frac{C_{f}}{\mathcal{L}},
  \nonumber
  \\
  &&C_{f}=-\frac{\hbar \kappa _{7}}{2}{\left (
    \frac{\ell _{P}}{\mathcal{L}}\right )}^{\Upsilon _{f}}\ell _{P}^{2}p^{2}.
\end{eqnarray}
The subscript $\pm{}$ in Eq.~{\eqref{fermiondispersion}} indicates the helicity-dependence
of the dispersion relation. Then the speed of massless fermions are~\cite{Alfaro:2002xz}
%
%e16 #&#
\begin{eqnarray}
  v_{\pm}^{f} (p,\mathcal{L})&&\simeq 1-\frac{3}{2}\kappa _{3}{(\ell _{P}
    p)}^{2}+{\left (\frac{\ell _{P}}{\mathcal{L}}\right )}^{\Upsilon _{f}+1}
  \nonumber
  \\
  &&{}\times \left [\kappa _{1}\mp \frac{\kappa _{7}}{2}(\ell _{P} p)
    \right ]+\kappa _{2}{\left (\frac{\ell _{P}}{\mathcal{L}}\right )}^{2
    \Upsilon _{f}+2}.
  \label{fermionspeed}
\end{eqnarray}
We should point out that Eq.~{\eqref{fermionspeed}} is obtained by expanding
to the first several orders of
${(\ell _{P}/\mathcal{L})}^{\Upsilon _{f}}$, so that this expansion might
not be valid for certain values of $\Upsilon _{f}$. Nevertheless we leave
this problem to be addressed after fixing the parameters in Eq.~{\eqref{fermionspeed}}
utilizing the aforementioned phenomenological picture.

In a very similar manner, we first drop the $\ell _{P}^{2}$ term in Eq.~{\eqref{fermionspeed}}
and adopt the mobile scale $\mathcal{L}=p^{-1}$ again, obtaining
%
%e17 #&#
\begin{eqnarray}
  v_{\pm}^{f} (p)&&\simeq 1+{\left (\ell _{P} p\right )}^{\Upsilon _{f}+1}
  \nonumber
  \\
  &&{}\times \left [\kappa _{1}\mp \frac{\kappa _{7}}{2}(\ell _{P} p)
    \right ]+\kappa _{2}{\left (\ell _{P} p\right )}^{2\Upsilon _{f}+2},
  \label{fermionspeed2}
\end{eqnarray}
which is valid for Majorana neutrinos. Obviously if we want both the signs
$\pm{}$ can be taken for the linear corrections, we must set
$\Upsilon _{f}=-1$ and obtain
%
%e18 #&#
\begin{equation}
  v_{\pm}^{f} (p)\simeq 1+(\kappa _{1}+\kappa _{2})\mp
  \frac{\kappa _{7}}{2}(\ell _{P} p).
  \label{fermionspeed3}
\end{equation}
However if in the limit $p\to 0$ we also require $v^{f}(0)=1$,
$\kappa _{1}+\kappa _{2}$ has to be zero and eventually we have
%
%e19 #&#
\begin{equation}
  v_{\pm}^{f} (p)\simeq 1\mp \frac{\kappa _{7}}{2}(\ell _{P} p),
  \label{fermionspeed4}
\end{equation}
with the parameters determined as
%
%e20 #&#
\begin{equation}
  \kappa _{1}+\kappa _{2}=0,\ \frac{2}{\kappa _{7}\ell _{P}}\approx 6.5
  \times 10^{17}~\text{GeV}{},\ \text{and }\Upsilon _{f}=-1,
  \label{neutrinoparameters}
\end{equation}
from which we obtain $\lvert \kappa _{7}\rvert \approx 37.5$, and it is
noteworthy that $\lvert \theta _{7}/\kappa _{7}\rvert \approx 1$.

After analyzing Majorana neutrinos, we also study the Dirac case in the
following for completeness. Hereafter we further assume that
$\kappa _{3}=0$ and the desired Dirac equations are
%
%e21 #&#
\begin{eqnarray}
  &&\left (i\frac{\partial}{\partial t}-i\vec{\sigma}\cdot \grad +
  \frac{\hat{C}}{2\mathcal{L}}\right )\xi (t,\vec{x})=0,
  \nonumber
  \\
  &&\left (i\frac{\partial}{\partial t}+i\vec{\sigma}\cdot \grad -
  \frac{\hat{C}}{2\mathcal{L}}\right )\chi (t,\vec{x})=0,
  \label{dirac2}
\end{eqnarray}
with $\chi =i\sigma _{2}\xi ^{*}$. Let us assume that $\chi{}$ is independent
of $\xi{}$ temporarily, thus a general Dirac spinor can be written as
%
%e22 #&#
\begin{equation}
  \Psi =
  \begin{bmatrix}
    \xi
    \\
    \chi
  \end{bmatrix}
  ,
\end{equation}
and then it is not hard to verify that the following equation reproduces
Eq.~{\eqref{dirac2}}:
%
%e23 #&#
\begin{equation}
  \left (i\gamma ^{\mu}\partial _{\mu}-\frac{\hat{C}}{2\mathcal{L}}
  \gamma ^{0}\gamma _{5}-m\right )\Psi =0,
  \label{dirac3}
\end{equation}
where the second term in the parenthesis explicitly violates Lorentz invariance
and we restore the mass term for later convenience. Of course Eq.~{\eqref{dirac3}}
may not be unique, but we only focus on it since we just attempt to show
that the results we just obtained are likely to apply to Dirac neutrinos
as well. To get the modified dispersion relation, we first multiply both
sides of this equation by
$(i\slashed{\partial}-\hat{C}\gamma ^{0}\gamma _{5}/2\mathcal{L}+m)$, obtaining
%
%e24 #&#
\begin{equation}
  \left (-\partial ^{2}-i\frac{\hat{C}}{2\mathcal{L}}\gamma _{5}\left (
  \gamma ^{\mu}\gamma ^{0}-\gamma ^{0}\gamma ^{\mu}\right )\partial _{
    \mu}-\frac{\hat{C}^{2}}{4\mathcal{L}^{2}}-m^{2}\right )\Psi =0,
\end{equation}
which, by straightforward calculation, can be simplified to be
%
%e25 #&#
\begin{equation}
  \left (-\partial ^{2}+i\frac{\hat{C}}{\mathcal{L}}\vec{\Sigma}\cdot
  \grad -\frac{\hat{C}^{2}}{4\mathcal{L}^{2}}-m^{2}\right )\Psi =0,
  \label{dirac4}
\end{equation}
with $\vec{\Sigma}:=\vec{\sigma}\otimes \text{diag}\{1,1\} $. Next we apply
$(-\partial ^{2}-i\hat{C}\vec{\Sigma}\cdot \grad /\mathcal{L}-\hat{C}^{2}/4
  \mathcal{L}^{2}-m^{2})$ to Eq.~{\eqref{dirac4}} and obtain the Klein-Gordon-type
expression
%
%e26 #&#
\begin{equation}
  \left ({\left (-\partial ^{2}-\frac{\hat{C}^{2}}{4\mathcal{L}^{2}}-m^{2}
    \right )}^{2}+\frac{\hat{C}^{2}}{\mathcal{L}^{2}}\grad ^{2}\right )
  \Psi =0,
  \label{dirac5}
\end{equation}
which should be satisfied by all components of the Dirac spinor. Ignoring
the mass term again and substituting a plane wave solution into Eq.~{\eqref{dirac5}},
we then read off the modified dispersion relation for Dirac spinors:
%
%e27 #&#
\begin{equation}
  E=\sqrt{p^{2}\pm \frac{\ell _{P}^{2}\kappa _{7}}{2\mathcal{L}}\left (
  \frac{\ell _{P}}{\mathcal{L}}\right )^{\Upsilon _{f}}p^{3}-
  \frac{\ell _{P}^{4}\kappa _{7}^{2}}{16\mathcal{L}^{2}}\left (
  \frac{\ell _{P}}{\mathcal{L}}\right )^{\Upsilon _{f}}p^{4}},
\end{equation}
or to the leading order:
%
%e28 #&#
\begin{equation}
  E\simeq p\pm \frac{\ell _{P}^{2}\kappa _{7}}{4\mathcal{L}}\left (
  \frac{\ell _{P}}{\mathcal{L}}\right )^{\Upsilon _{f}}p^{2}.
  \label{diracmdr}
\end{equation}
It is obvious that Eq.~{\eqref{diracmdr}} also leads to Eq.~{\eqref{fermionspeed4}}
under the similar assumptions:
%
%e29 #&#
\begin{equation}
  v^{f}=\frac{\partial E}{\partial p}\simeq 1\pm \frac{\kappa _{7}}{2}(
  \ell _{P} p).
\end{equation}
As a result, no matter whether neutrinos are Dirac particles or Majorana
particles, the phenomenological constraints from cosmic neutrino data analyses~\cite{Amelino-Camelia:2016ohi,Huang:2018ham,Huang2019,Huang2022}
can always be satisfied with the same parameters.

Most importantly, we further analyze the two parameters
$\Upsilon _{\gamma}$ and $\Upsilon _{f}$ which play a crucial role in producing
the desired energy dependence and helicity dependence. Let
$A^{i}_{a}$ be the real connection variables, then these two parameters
are determined by~\cite{Alfaro2002a,Alfaro:2002xz}
%
%e30 #&#
\begin{eqnarray}
  &&\bra{W,\uline{\vec{E}},\uline{\vec{B}}}\dots A_{ia}\dots
  \ket{W,\uline{\vec{E}},\uline{\vec{B}}}\approx \dots
  \frac{1}{\mathcal{L}}\left (\frac{\ell _{P}}{\mathcal{L}}\right )^{
  \Upsilon _{\gamma}}\dots ,
  \nonumber
  \\
  &&\bra{W,\xi}\dots A_{ia}\dots \ket{W,\xi}\approx \dots
  \frac{1}{\mathcal{L}}\left (\frac{\ell _{P}}{\mathcal{L}}\right )^{
  \Upsilon _{f}}\delta _{ia}\dots
  \label{upsilon}
\end{eqnarray}
respectively, where $\ket{W,\uline{\vec{E}},\uline{\vec{B}}}$ and
$\ket{W,\xi}$ are the corresponding WBSCs. A straightforward and satisfactory
choice for $\Upsilon _{\gamma}$ and $\Upsilon _{f}$ may be
$\Upsilon _{\gamma}=\Upsilon _{f}=1$~\cite{Alfaro2002a,Alfaro:2002xz} since
it makes both the expectation values in Eq.~{\eqref{upsilon}} vanish in the
limit $\hbar \to 0$ and a vanishing expectation value of this type means
that the spacetime discreteness disappears once the quantum effects are
turned off. Thus it seems that $\Upsilon _{\gamma}=-1/2$ and
$\Upsilon _{f}=-1$ are both unacceptable. However if we speculate that
$\Upsilon _{\bullet}$~(the slot $(\bullet )$ represents $\gamma{}$ or
$f$) is a function of the spin of the corresponding particles,\footnote{It
  may also depend on other quantum numbers, but here we have few clues what
  the form the dependence could be, so we take them into consideration by
  introducing some parameters.} this problem might be addressed as follows.
For convenience, we assume that the limit $\hbar \to 0$ can be taken continuously
and one permissible way to take this limit is
$\hbar (t)=\hbar _{0}(1-t)$ where $0\le t\le 1$ and $\hbar _{0}$ is the
Planck constant usually measured in laboratories. We further rewrite the
spins of photons and fermions as $s(t;\gamma )=\hbar (t)/\hbar _{0}$ and
$s(t;f)=\hbar (t)/2\hbar _{0}$ respectively such that the spins vanish
when $\hbar{}$ is zero and take their ordinary values when $\hbar{}$ is
the real-world one. There are still many possibilities of how
$\Upsilon _{\bullet}{}$ depends on the spin $s$, while here as an example
we choose the simplest one that $\Upsilon _{\bullet}{}$ depends on
$s$ linearly: $\Upsilon _{\bullet}(s)=s-3/2$. Then the expectation values
behave like
%
%e31 #&#
\begin{eqnarray}
  &&\frac{1}{\mathcal{L}}\left (\frac{\ell _{P}}{\mathcal{L}}\right )^{
  \Upsilon _{\gamma}(s)}\propto \sqrt{1-t}^{-t-\frac{1}{2}},
  \nonumber
  \\
  &&\frac{1}{\mathcal{L}}\left (\frac{\ell _{P}}{\mathcal{L}}\right )^{
  \Upsilon _{f}(s)}\propto \sqrt{1-t}^{-\frac{t}{2}-1},
\end{eqnarray}
which diverge as $t\to 1$ so the linear form is not valid. In addition
there is no free parameter describing the effects of other quantum numbers
and this also makes the linear form disfavored. We then try a quadratic
function: $\Upsilon _{\bullet}(s)=a_{1}s^{2}+a_{2}s+a_{3}$ where
$a_{i}$ with $i=1,2,3$ are parameters that may be dependent on other quantum
numbers. A direct calculation suggests that
%
%e32 #&#
\begin{equation}
  \Upsilon _{\bullet}(s)=(3+2a_{3})s^{2}-(\frac{7}{2}+3a_{3})s+a_{3}.
\end{equation}
We do not have any knowledge of the exact value of $a_{3}$, but let us
consider the expectation values:
%
%e33 #&#
\begin{eqnarray}
  &&\frac{1}{\mathcal{L}}\left (\frac{\ell _{P}}{\mathcal{L}}\right )^{
  \Upsilon _{\gamma}(s)}\propto \sqrt{1-t}^{(3+2a_{3})t^{2}-(
  \frac{5}{2}+a_{3})t-\frac{1}{2}},
  \nonumber
  \\
  &&\frac{1}{\mathcal{L}}\left (\frac{\ell _{P}}{\mathcal{L}}\right )^{
  \Upsilon _{f}(s)}\propto \sqrt{1-t}^{(\frac{3}{4}+\frac{a_{3}}{2})t^{2}+(
  \frac{1}{4}+\frac{a_{3}}{2})t-1},
\end{eqnarray}
which are zero at $t\to 1$ so long as $a_{3}>0$ and recover the desired
behavior at $t\to 0$. The parameter $a_{3}$ is likely to encode the potential
effects of other quantum numbers as we expect and as a result the problem
of turning off quantum effects is likely to be solved. Another associated
problem is whether the perturbative expressions Eq.~{\eqref{photonspeed}}
and Eq.~{\eqref{fermionspeed}} are still valid with the present parameter
choice. It is clear that for photons, since
$2+2\Upsilon _{\gamma}=1>0$, the expansion with respect to small
$\ell _{P}/\mathcal{L}$~(or $\ell _{P} k$) does not bring any new difficulty.
In the fermion case, instead of perturbing with respect to
$\ell _{P}/\mathcal{L}$, one can alternatively expand with respect to
$\ell _{P}$. With this fact we can easily see that the dispersion relation~{\eqref{fermiondispersion}}
again yields the speed~{\eqref{fermionspeed}}. As a result, although the
choices of parameters may violate the perturbative expansions~{\eqref{photonspeed}}
and~{\eqref{fermionspeed}}, the phenomenologically determined ones are still
consistent with these perturbative expressions and the associated analyses
are valid as well.

Lastly, we should point out that to obtain the desired dispersion relations
Eq.~{\eqref{photondispersion}} and Eq.~{\eqref{fermiondispersion}}, we need
to introduce the description of LQG based on weave states~(WBSCs in this
work)~\cite{Alfaro:2002xz,Alfaro2002a}, and consequently the results depend
on the assumption of weave states and also the assumed properties of weave
states. LQG still is an incomplete theory (in the literal sense that its
``Hamiltonian constraint'' has not been solved) and therefore all the predictions
obtained from LQG require some assumptions. It is of notice that the much-studied
description of LQG based on weave states~\cite{Alfaro:2002xz,Alfaro2002a}
has the sort of phenomenological implications described in this manuscript.
Therefore the results from LQG with assumed weave states make sense because
they can be directly related to observable phenomena and conversely the
phenomenological analyses may be used to examine the description of LQG
based on weave states, leading towards better understanding of the LQG
framework and further progress of LQG theories.

In summary, we find that loop quantum gravity~(LQG) provides a viable way
to understanding the phenomenological suggestion of the speed variations
of cosmic photons and neutrinos. Comparing the theoretical calculation
and phenomenological picture we determine several parameters of LQG induced
LIV, which conversely may be helpful for the understanding of LQG itself.
For example, we determine the exact values of $\Upsilon _{\gamma}$ and
$\Upsilon _{f}$, which reflect some properties of the WBSCs. Besides, we
find that $\theta _{7}$ and $\kappa _{7}$ are of the desired order of magnitude
and a simple assumption of $\theta _{7}\approx \kappa _{7}$ reproduces
the phenomenological result
$E_{\text{LIV}}^{\gamma }/E_{\text{LIV}}^{\nu }\approx 1:2$, which is an
interesting observation. We thus conclude that LQG seems to be a promising
theory for explaining the novel picture suggested from phenomenological
analyses on the time of flight of cosmic photons~\cite{Shao2010f,Zhang2015,Xu2016a,Xu2016,Amelino-Camelia:2016ohi,Amelino-Camelia:2017zva,Xu2018,Liu2018,Li2020,Zhu2021a,Chen2021}
and neutrinos~\cite{Amelino-Camelia:2015nqa,Amelino-Camelia:2016fuh,Amelino-Camelia:2016ohi,Amelino-Camelia:2017zva,Huang:2018ham,Huang2019,Huang2022}.
With the suggested LQG parameters, we expect that more observable signals
could testify the predictions of loop quantum gravity in the future.

%\section*{CRediT authorship contribution statement}

% \begin{conflict}
%   The authors declare that they have no known competing financial
%   interests or personal relationships that could have appeared to influence
%   the work reported in this paper.
% \end{conflict}
%
%\begin{conflict} % None declared.
%\end{conflict}

% \begin{dataavailability}[title={Data availability}]
%   No data was used for the research described in the article.
% \end{dataavailability}
%
%\begin{dataavailability}[title=Data availability]
%\end{dataavailability}

\begin{acknowledgments}
  This work is supported by National Nature Science Foundation of China~(Grant
  No.~12075003).
\end{acknowledgments}

\appendix
\section{Conventions\label{conventions}}
  %\def\the...{}
  %\reset{}{}
  %\appendix{}
  %\appendix*{}
  %\appendix{Conventions}
  %%LEAP%%%\label{appA}
  %\label{conventions}

  The signature is chosen to be $\eta ^{\mu \nu}=(+1,-1,-1,-1)$. The Pauli
  matrices $\sigma ^{\mu}{}$ are
  %
  %eA1 #&#
  \begin{eqnarray}
    && \sigma ^{0}=\left (
    \begin{array}{c@{\quad}c}
        1 & 0
        \\
        0 & 1
      \end{array}
    \right ),\ \sigma ^{1}=\left (
    \begin{array}{c@{\quad}c}
        0 & 1
        \\
        1 & 0
      \end{array}
    \right ),
    \nonumber
    {}
    \\
    && \sigma ^{2}=\left (
    \begin{array}{c@{\quad}c}
        0 & -i
        \\
        i & 0
      \end{array}
    \right ),\ \sigma ^{3}=\left (
    \begin{array}{c@{\quad}c}
        1 & 0
        \\
        0 & -1
      \end{array}
    \right ),
    \label{sigma}
  \end{eqnarray}%
  and $\bar\sigma ^{\mu}:=(\sigma ^{0},-\vec{\sigma})$. We adopt the same
  convention for the $\gamma{}$ matrices as that in the book of Peskin and
  Schroeder~\cite{Peskin:1995ev}, and hence the $\gamma{}$ matrices are expressed
  as
  %
  %eA2 #&#
  \begin{equation}
    \gamma ^{\mu}=\left (
    \begin{array}{c@{\quad}c}
        0                  & \sigma ^{\mu }
        \\
        \bar\sigma ^{\mu } & 0
      \end{array}
    \right ).
  \end{equation}
  Correspondingly,
  $\gamma _{5}:=i\gamma _{0}\gamma _{1}\gamma _{2}\gamma _{3}$, which takes
  the simple form: $\gamma _{5}=\text{diag}\{-1,-1,+1,+1\}{}$, and the helicity
  projection operators are
  %
  %eA3 #&#
  \begin{equation}
    P_{L}=\frac{1-\gamma _{5}}{2},\quad P_{R}=\frac{1+\gamma _{5}}{2}.
  \end{equation}

%\bibliography{ref-lqgphe}

\end{document}